# An architecture for "Web Of Things" using SOCKS protocol based IPv6/IPv4 gatewaying for heterogeneous communication.

P. Shrinivasan. R. Patnaikuni
Walchand Institute Of Technology,
Solapur University, Solapur, India.

Raj. B. Kulkarni.
Walchand Institute Of Technology,
Solapur University, Solapur, India.

ABSTRACT

"Web Of Things" evolved from "Internet Of Things". Lot of research has been done in designing architecture for "Web Of Things". Two main architectures are Smart gateway based architecture and embedded Web Server based architecture. These architectures address some of the basic and essential issues relating to Service Oriented Architecture for "Web Of Things". Taking into consideration the period of coexistence of IPv4 and IPv6 we propose an architecture using SOCKS protocol based IPv6/IPv4 gatewaying and refinements which facilitates smooth heterogeneous communications between the IPv6 and IPv4 enabled embedded nodes and can potentially be used to prevent security threats like Denial-of-Service (DoS) attacks on embedded devices attached to the web and increase its performance. Our architecture provides a way for caching responses from device and thereby increasing its efficiency and performance and yielding quick response times.

1 INTRODUCTION

Today, the Web is a global platform for information based applications, but that is about to change. What is driving this is the rapidly changing incremental cost of networking for all kinds of devices. It is now easy to integrate radio-frequency components alongside digital circuitry for microcontrollers. We are in the midst of a proliferation of devices that are largely invisible as they are embedded within everyday objects from toasters to cameras and cars. Microcontrollers are the fastest growing segment of the computer industry, with hundreds in every home. These devices are programmed to serve a single purpose, and today are mostly isolated. Networking them will allow many new kinds of applications that add values in the way that the original manufacturer may not have envisaged. In this scenario there is strict need for good scalable and reliable architecture for existing "Web Of Things". Making the smart things interconnectable such that bits can be transferred between devices is only the first step, more works are expected to make smart things interoperable such that they are understandable with each other. Interoperability is particularly essential, and a must, to build system with various devices. In this paper we will discuss some of the possible loop holes in existing architecture and propose solutions towards better architecture for "Web Of Things".

2. RELATED WORK

During the early stages of "Web Of Things" [10] two architectures where proposed these architectures rely on concept that sensors act as a RESTful resources [7]. Here sensors can be of any type. Main architecture is REST based architecture [5]. This allows the end devices to be accessible through HTTP protocol using RESTful APIs [9]. The two architectures are Web oriented architectures. Creating resource oriented architecture has been the main achievement of "Web Of Things". The first architecture earlier proposed [1] is for direct access to the API on devices which have capability to run embedded web servers on them hence the capability to interact using REST principles. Second architecture has been for access to API on smart gateways which act like an intermediately in between end devices and web server [6]. Even earlier similar architecture was proposed [2][9] but they are not a promising one. Ostermaier et al. [11] presented a prototype using programmable low-power WiFi modules for connecting things directly to the web. They leverage the ubiquity of IEEE 802.11 access points and the interoperability of the HTTP protocol. Using a loosely coupled approach, they enable seamless association of sensors, actuators, and everyday objects with each other and with the Web. All those works demonstrate convincingly that it is possible to integrate smart things directly into the Web now. Detailed description about REST principles and Resource Oriented Architecture can be found at [3].

3 EXISTING ARCHITECTURES AND METHODS

Offering direct access to devices is limited to fact that they should have embedded web server though there is steady increase in use and manufacturing such devices. Such Web enabled devices can be directly integrated and make their RESTful APIs directly accessible on the web. This integration process is shown in Figure 1. Each device has an IP address and runs a web server on top of which it offers a RESTful API to the mashup developer.

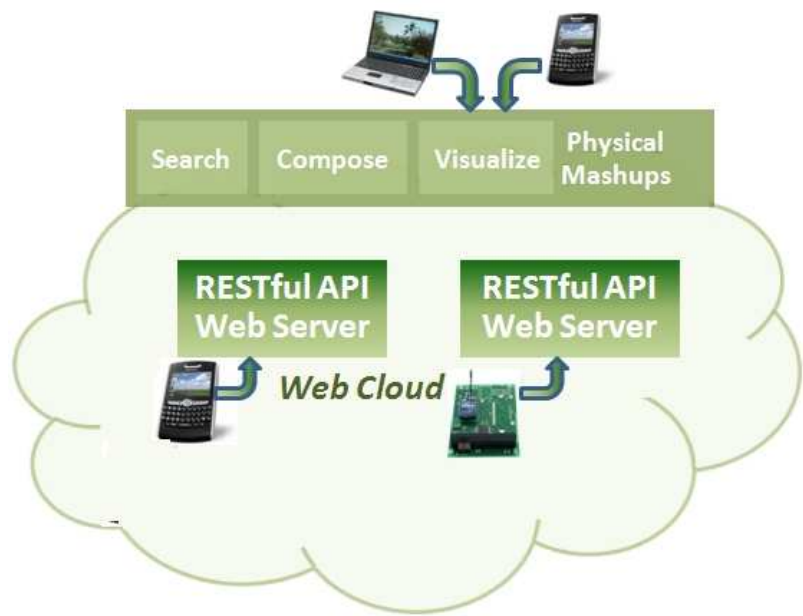

Figure 1: Architecture for direct access through API for IP enabled embedded Devices [1].

The second architecture which is based on having smart gateway, providing interface between devices which do not have embedded IP but are capable of interacting in own custom protocols, example of such type of devices are zigbee[4]. Here each smart gateway features a web server equipped with ability to interact with non IP based end devices [8]. The web server in smart gateway is one which provides access to the devices. As an example, consider a request to a sensor node coming from the web through the RESTful API. The gateway maps this request to a request in the proprietary API of the node and transmits it using the communication protocol which the sensor node understands. The best advantage of using smart gateways is that it can support multiple types of devices using proprietary protocols for communication as shown in Figure 2.

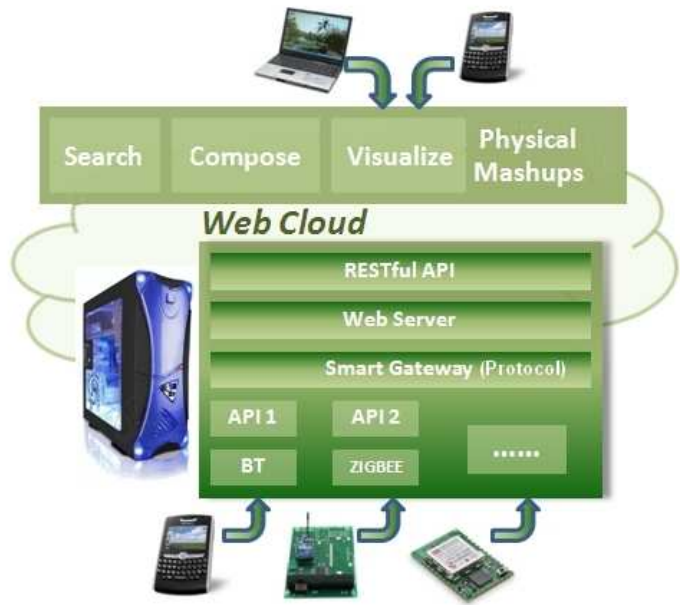

Figure 2: Smart gateway based architecture [1].

4. NEED FOR REFINEMENTS

Many Large-scale enterprise automation, metering and monitoring systems etc require end-to-end addressing, security, mobility, traffic multiplexing, reusability, maintainability, and webservices which are globally scalable. With the IETF's long-lived, standards. IPv6 [14] based 6LowPAN [15] stands out as promising option for "Web Of Things". Hence we focus on architecture involving IP enabled devices. Standalone isolated wireless IP enabled embedded networks are not viable for applications that require accessing services in the Web. Wireless IP enabled embedded networks are more vulnerable to misuse than wired networks.

In addition, a malicious device or malfunctioning device may be present in the network. It can analyze the communication in the network and do several attacks by sending invalid data to another device, or it may create scenarios like Denial-of-Service (DoS) attacks. In particular, it can even block all communication by constantly interfere the transmission.

Though the existing architectures address basic issues they do not deal with heterogeneous IP enabled embedded devices here by word heterogeneous we mean nodes supporting IPv4 or

IPv6. The following stated points prompt us to go for further refinements in existing architecture.

- When some nodes are IPv6 enabled and some are not the communication between them poses some issues.
- In resource oriented web of things if IPv6 node requests for IPv4 node or IPv4 node request for IPv6 then there has to be proper gatewaying of communication.
- Moreover embedded devices have very less amount of hardware resources. They work on very low power and this puts a serious limitation on applications running on them E.g. Web server embedded on device cannot function as robustly as standalone web server running on highly powerful computer having good computing resources.
- Embedded web servers cannot handle burst of service requests due to their limited resources and computational ability.

The above limitations prompted us to go for better, refined and robust REST based architecture.

A promising approach which could tackle these situations uses one or more devices in the wireless IP enabled embedded network as a gateway to an external network. This external network is the Internet in case of "Web Of Things". The Gateway provides a connection to another network and all inter and intra network communications are controlled and monitored. Consequently, the devices must be able to communicate when the gateway is available and when it is unavailable. And further optimization of JSON messages used for application level communication leads to better performance and results.

5. NEW ARCHITECTURE AND REFINEMENTS

The new architecture uses gateway web server as in traditional networks. The gateway web server here is typically a dual IP stack machine using SOCKS [12] protocol based Gateway Mechanism for communication between IPv4/IPv6 nodes [13]. The gateway web server manages the transition from IPv4/ IPv6 link by relaying two "terminated" IPv4 and IPv6 connections at the "application layer". Since the transition is SOCKS protocol based transition and also no new protocols are introduced the communication semantics are preserved in heterogeneous environment of IP v4 and IPv6. The whole process and mechanism is detailed in RFC 3089 [13], we just focus on the applicability of the SOCKS protocol based IPv4-IPv6 gatewaying mechanism in "Web Of Things" scenario. The Figure 3 shows the overview.

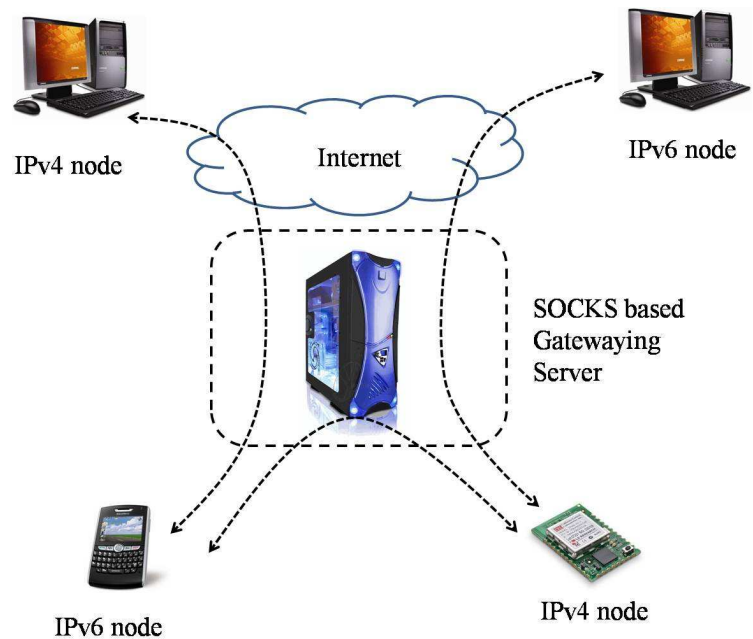

Figure 3: Overview of proposed mechanism of Gatewaying in "Web Of Things"

The new architecture also uses dictionary type mappings to reduce overall length of JSON messages used in Service Oriented Architecture of "Web Of Things". Its gateway web server is computationally powerful enough to store dictionary of mappings and tackle security threats like Denial-of-Service (DoS) attacks on devices. Gateway web server allows monitored access to the devices and this gateway web server is introduced in both the architectures discussed in section 3. The functions of gateway web server apart from acting as a gateway for the communication in between IPv4 and IPv6 enabled embedded nodes are.

- To receive request from end users from internet web, request may be from IPv4 or IPv6 enabled network.
- To check/block the request if suspicious behavior in request pattern is observed.
- Caching responses from devices and avoiding unnecessary repeat requests.
- To respond with proper status if devices has stopped working.

- To forward and to act as gateway for IPv4/IPv6 based request to APIs on devices in coded format and decode JSON response from devices and construct meaningful response for the end user.

By coding and decoding we mean that, suppose originally an API on embedded devices (E.g. power sensors) have been programmed to return JSON response for respective JSON request as shown below.

JSON Request:

{

"values":[ {"NoOfDevces":[2]},]

}

JSON Response:

[

{

"deviceName": "ComputerAndScreen",

"currentWatts": 50.52,

"KWh": 5.835,

"maxWattage": 100.56

},

{

"deviceName": "Fridge",

"currentWatts": 86.28.,

"KWh": 4.421,

"maxWattage": 288.92

},

{...}

]

In coding process does the task of replacing the variable names in JSON request by short codes.

- Example  "NoOfDevices" by "ND"

In decoding process task of replacing short codes generated in coding process back into original variable names is done so that a real world meaningful response can be returned to end user.

- Example "ND" back by "NoOfDevices".

Suppose if we use following set of mappings for coding and decoding JSON objects variable names.

- NoOfDevices ---ND.
- deviceName  --- DN.
- currentWatts ---CW.
- maxWattage ---MW.

The new JSON Request and JSON Response messages structures will be like:

JSON Request:

{

"values":[ {" ND ":[2]},]

}

JSON Response:

[

{

"DN": "ComputerAndScreen",

"CW": 50.52,

"KWh": 5.835,

"MW": 100.56

},

{

"DN": "Fridge",

"CW": 86.28.,

```
"KWh": 4.421,
"MW": 288.92
},
{...}
]
```

The RESTful APIs on embedded devices are designed to understand and processes coded request and generate coded responses. The gateway web server manages coding and decoding for end user. This coding and decoding method reduces length of request and response messages generated between gateway web server and physical devices resulting in quicker response times by physical embedded devices.

The gateway web server also serves the function of caching responses from devices thereby avoiding excessive load on embedded devices. The caching feature also ensures that embedded devices can enter into power save mode. The gateway web server can be used to monitor request pattern and block Denial-of-Service (DoS) attacks.

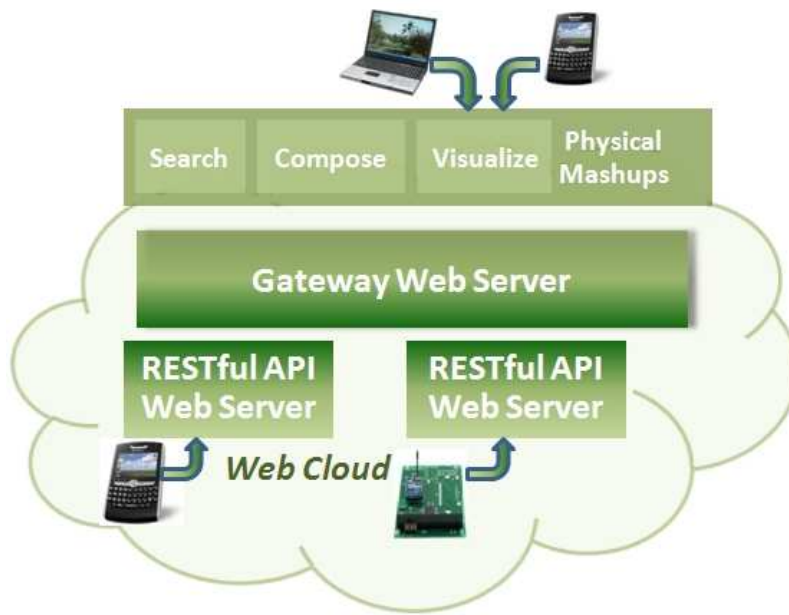

Figure 4: Direct access API architecture with gateway web server.

In case of architecture using smart gateways for proprietary and custom protocols of devices the web server itself acts like a gateway web server, as show in Figure 5.

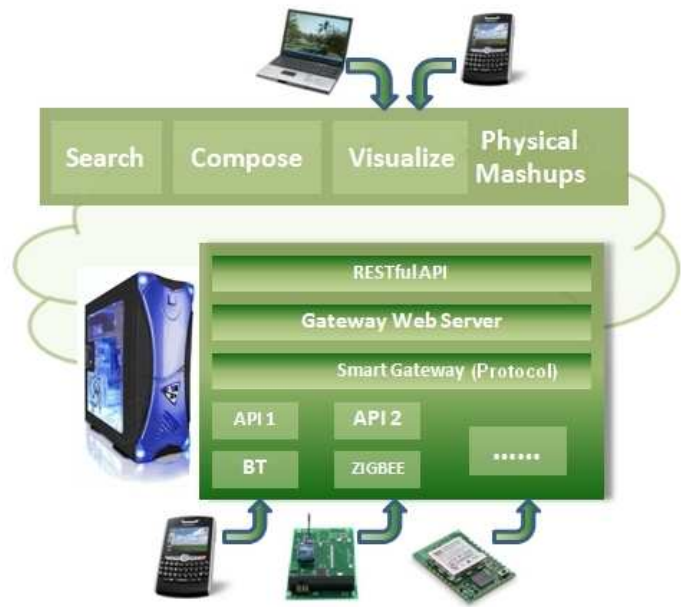

Figure 5: Location of gateway web server in case of smart proprietary protocols gateway architecture.

6. IMPLEMENTATION AND CONCLUSION

These proposed refinements for existing architecture have been tested on with a open source emulator Qemu [16] emulating ARM architecture and running miniature web server responding to HTTP requests with JSON messages with gateway web server implemented on Apache Tomcat and results have been promising, showing increase in performance based on average response time. Our next task is to actually implement our proposed architecture with embedded devices preferably ARM based devices supporting IPv6 and with gateway web server using SOCKS protocol based gatewaying mechanism for communication between IPv4/IPv6 nodes. A Java based SOCKS implementation is being tried out for possible integration with Apache Tomcat. Integration of security features suitable to low powered embedded into the proposed architecture is a future objective.